\documentclass[]{article}
\usepackage{times,epsfig}
\usepackage{makeidx,german}
\title{Business in the Grid}
\author{%
Erich Schikuta{\small $~^{\#}$} \and Thomas Weish\"aupl{\small $~^{\#}$} \and Flavia Donno{\small $~^{+}$} \and Heinz Stockinger{\small $~^{*}$} \and
Elisabeth Vinek{\small $~^{+}$} \and Helmut Wanek{\small $~^{\#}$} \and Christoph Witzany{\small $~^{\#}$} \and Irfan Ul Haq{\small $~^{\#}$}
\vspace{1.6mm}\\
\fontsize{10}{10}\selectfont\itshape
$~^{\#}$ University of Vienna, Department of Knowledge and Business Engineering \\
\fontsize{10}{10}\selectfont\itshape Rathausstra"se 19/9, A-1010 Vienna, Austria \\
\fontsize{10}{10}\selectfont\itshape $~^{*}$Swiss Institute of Bioinformatics, CH-1015 Lausanne, Switzerland\\
\fontsize{10}{10}\selectfont\itshape $~^{+}$CERN, CH-1211 Geneve 23, Switzerland
}

\date{}

\begin{document}
\maketitle

\section*{Abstract}
From 2004 to 2007 the Business In the Grid (BIG) project took place
and was driven by the following goals: Firstly, make business aware of
Grid technology and,  secondly, try to explore new business models. We
disseminated Grid computing by mainly concentrating on the central
European market and interviewed several companies in order to gain
insights into the Grid acceptance in industrial environments. In this
article we present the results of the project, elaborate on a
critical discussion on business adaptations, and describe a novel
dynamic authorization workflow for business processes in the Grid.

\section{Introduction}
\label{chap:intro}

Grid computing finds its origin in the academic world where it was
``created'' and has gained a lot of popularity in the last decade. Since then, many national and international Grid oriented solutions have been devised specially to harness huge data and solve 
computationally intensive problems. A few years ago, commercial
partners also joined Grid efforts in order to profit from this promising
technology. However, the Grid community is still in search of the
so called ``killer application'' and is experimenting with application scenarios where the
commercial world can really make profit by utilising Grid technologies.


Nowadays it can be learned from the developments in the IT market that there is an obvious shift in the
source of profits, from sales of products to the provision of on-demand services \cite{ngg3}. This
leads to the development of novel business models where new application domains will share applications and
resources. Thus for companies in the IT market to be successful the availability of techniques
allowing for integration and collaboration of resources of any kind is a crucial issue.

The Grid promises the revolution of the Internet by a novel and advanced support for collaboration
providing homogeneous access to virtual resources without revealing the heterogeneous manner of the
underlying real world. Basically the Grid resembles a distributed computing model supporting the
selection, sharing, and aggregation of geographically distributed \hbox{''}autonomous\hbox{``}
resources dynamically at runtime depending on their availability, capabilities, performance, costs,
and users' quality-of service requirements via Grid-enabled Web services \cite{gridbus07}.

In the literature a large number of different definitions of the notion of Grid can be found \cite{Stocki07}.
This is connected with an ongoing discussion within the research community what the specific
characteristics of a Grid really are and what makes it so different from the Internet and the Web.
From our business oriented point of view we identified three necessary specifics of the Grid infrastructure to allow for business workflows, which are:

\begin{itemize}
  \item \textbf{Transparency.} The parties involved in sharing of resources are anonymous to each other, that means the consumer of a resource does not need any knowledge about the provider of the resource and vice versa.
  \item \textbf{Generalized Quality of Service.} QoS is more than technical properties. It is better described by the notion of SLA (Service Level Agreement) and has to comprise all necessary aspects of business resulting in a \hbox{''}trust relationship\hbox{``} between customer and provider. This can be confidentiality, data integrity, non-repudiation, accountability, \ldots
  \item \textbf{Brokerage.} Resources are made available by a Broker, which is in our terminology a policy-based mediator of services and business workflows aiming for maximizing profit of all partners.
\end{itemize}

In our vision it will be possible to sell software and resources as a service and not as a good in
the near future.

For example, \emph{"Writing a letter"} can be as simple as using a telephone: Forget buying software and hardware!
All we need is a simple interface to the services on the Grid, which could be a cell phone. Both the wordprocessor
functionality and the necessary physical resources (processor cycles
and storage space) are available as services on the Grid; and we pay just for the usage transparently via our telephone bill.


Until now the research community mainly focused on the technical aspects of Grid computing
neglecting commercial issues. However now there is a focus shift towards the commercial
exploitation of Grid computing.


In 2004 we created the project Business in the Grid (BIG) in order to
analyse the IT market and its current acceptance to Grid computing in
business environments. We mainly focused on the central European
market but believe that the results are representative for other
countries, too.

Preliminary results of the projects have already been published
earlier (\cite{BIGResult}). In this article, we provide more in-depth
and updated information on our findings. The article is organised as
follows. We will first review the project goals in
Section \ref{sec:goals} and then revise existing business models. These
models are then used to analyse the IT market by first proposing a
dictionary (Section ~\ref{sec:dictionary}) and then a detailed market
analysis along with an on-line survey. Finally, we discuss new
business models and present as novel scientific result of the project
the gSET method for dynamic authorization in business workflows on the Grid.

%
%
\section{BIG Project Goals}
\label{sec:goals}

The project Business In the Grid (BIG) was created with the
purpose to analyse the IT market in order to find out if there is
a market potential to use Grid technologies in a commercial
environment. In detail, the project had the following goals, as
described in~\cite{WeisBIG05}:

\begin{enumerate}

\item {\bf Revisiting of existing (E-)business models for the Grid.}

Existing (E-)business models were revisited for a possible adaptation to the Grid.
Reasons for the failure of existing business models for E-Commerce in the Internet
and Business to Business (B2B) were analysed. It was observed that the Grid can provide
chances of success for these models by new \emph{transparent layers}
(included security mechanisms, support of dynamic services, etc.).

\item {\bf Market potential analysis and information dissemination.}

Information dissemination and market potential and demand analysis
was mainly focused on the Austrian economy landscape, but some other countries such as Germany, Czech Republic and USA were also kept in view.

\item {\bf Development of novel business models for the Grid.}

The main goal of the BIG project was to find novel business models
enabled by the Grid infrastructure. New ways of doing business in
the Grid were discovered. It was realized that there are possibilities for dynamic
collaborations, new project workflows, software on demand, dynamic
resource-management, resource on demand, application service
providers, and other Grid information society components.
                                                                                \end{enumerate}

\section{Revisiting of Existing Business Models}\label{chap:revisiting}

\subsection{Economy Grid Layers}

In the first phase of the BIG project, a broad state of the art analysis was done that revealed very interesting information. It was noticed that no standard terminology is used in ``economy Grid'' projects in the community. Until now no clear understanding and categorization exists in this area. Problem formulations of many ``economy Grid'' projects have difficult lingual descriptions, based on vague assumptions thus resulting in weak problem and requirement specifications.

The state of the art analysis shows that economic aspects influence Grid computing projects on different layers. The terms business, commerce, and economy are not used consistently. This makes it difficult to proceed efficiently to the commercialization of virtual resources.

The purpose of the following chapter is to clarify the relation
between Grid computing and economy with the final goal to transform
virtual resources to commodities. We proposed a layered architecture
~\cite{WeisBIG05} to describe different economic aspects of the Grid. The model is called Economy-Grid layer (EG layer) model.

\subsubsection{Methodology and Background}\label{egl-sec:methode}

The EG layer model is derived from the background of historical observation. Different hypotheses and theories \cite{Free90} exist which aim to explain the reasons for the technological developments in relation to economy and innovation.

\begin{figure}
  \centering
  \includegraphics[angle=0,width=0.8\columnwidth]{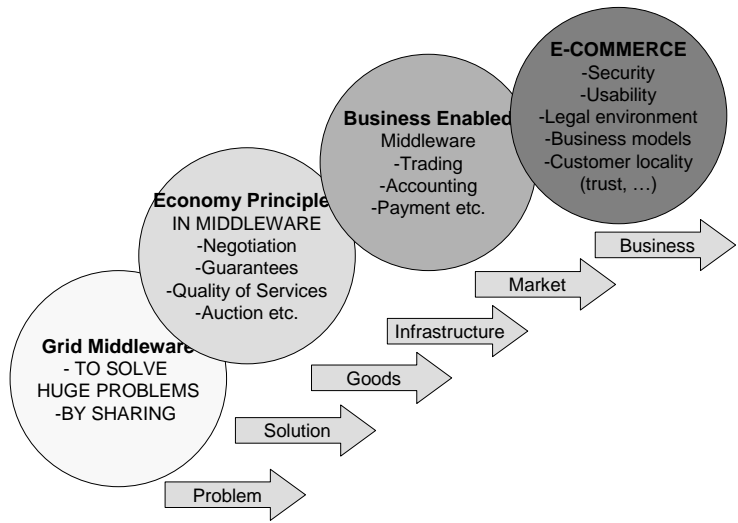}\\
  \caption{Grid development to business}\label{egl-fig:eg-development}
\end{figure}

We use simplified categories describing the history of the development of existing technologies. The categories result from a ``problem-solution'' process with the respective steps: problem, solution, good, infrastructure, market, business. Figure \ref{egl-fig:eg-development} visualizes the mentioned process by arrows below the cycles. We show the validity of the simplified process by two examples:

Telephony is a representative technology, which has followed this historical process \cite{Bell05,Fr95}. At the beginning the transmission of voice signals between two dislocated points was achieved (problem-solution). It was not trivial that this solution developed into a good. It took many years from the discovery of the telephone to the good, which means that somebody would buy and pay for the service or good. An infrastructure had to be built to deal with the good. Accounting and other systems were necessary. By the infrastructure it was possible that a market with customers could be
established. Initially, the telecom markets were quite small, homogeneous, and limited (e.g. one provider per country or region). Therefore, we distinguish a further step, called business, which represents a multi provider market.

As another example for this process the evolution and usage of the
Internet can be mentioned. Considering the Internet, which was
invented to solve communication challenges in the area of military
defence, is now a communication infrastructure for everybody, used
also for doing commerce (e.g. Amazon, EBay etc.). Obviously out of a problem and its solution arises a new good. This good can be traded in an adapted infrastructure, which creates a market and therefore a new commercial business.

The cycles of Figure \ref{egl-fig:eg-development} visualize a possible evolution of Grid technology. Grid middleware was developed to solve huge computational problems by sharing of resources inside a virtual organization (see \cite{FoKe99}). The Grid middleware uses not only principles and concepts of computer science, but also economy to provide the capabilities mentioned above, e.g. \cite{BeCaCaMiStZi03}. This middleware is a new commodity or product. Until now, the developments are in progress and no finalized ``business enabled'' middleware exists (see \cite{Globus}), which can establish a new market or field of commerce.

By the allocation of the Grid development to the historical process of common technologies we presume the future development and derive the EG layer model.  All projects described in the next section can be categorized in this way. However,  this list should not be taken as exhaustive.

\subsubsection{The Economy Grid Layer Model}\label{egl-sec:eglmodel}

The model results from the observations and conclusions mentioned above.
We propose the Economy-Grid layer (EG layer) model, to better
understand the problems in context of using a Grid infrastructure as a
new good or product.

The EG-layer model consists of four layers with the following characteristics:
\begin{description}
    \item[EG1 Integration Layer: Grid using economic principles]  ~

Economic principles, concepts, and experience are integrated into the Grid and influence
\emph{developments} of Grid infrastructure, e.g. resource usage can be optimized by the
adaptation of auction principles. Typical representatives are the
economy based replica optimizer~\cite{BeCaCaMiStZi03} and the Gridbus project \cite{Gridbus} with the
 GRACE (Nimrod-G) component \cite{Buy02}. The GEMSS project \cite{GEMSS} and the NeuroWeb project \cite{Schi02} are also examples
  for EG1, which negotiate quality of services (QoS) between clients and service
providers.

    \item[EG2 Commercialization Layer: Selling Grid software] ~

Companies create products or services by using Grid software or some Grid components for
``homogeneous'' organizations.  They \emph{sell} the recent open source software combined
with self-developed software modules and services. Representatives of EG2 are for example
 DataSynapse~\cite{DataSynapse}, Avaki~\cite{Avaki}, Univa~\cite{Univa} or
  Orcale.

    \item[EG3 Enabling Layer: Business enabled Grid] ~

The business enabled Grid establishes an open Grid, with similar properties as
the Internet for information today. An infrastructure for a market has to be
provided. In a Grid the resources can not be free, but accessible under user
constraints. A market can regulate resource sharing in a satisfying way for a
resource provider and customer. \emph{Single} companies need trading, accounting
and payment mechanisms. Representative components of EG3 are the BEinGrid
\cite{Beingrid}, Adaptive Service Grid (ASG) \cite{Asg}, CHALLENGERS
\cite{Challengers}, Financial Service Grid (FinGRID) \cite{Fingrid}, BREIN
\cite{Brein}, GridBank \cite{BaBu03}, GRIA \cite{GRIA}, an OGSA-Based Accounting
System \cite{SaGarEl04}, and the GGF GESA-WG \cite{GESA-WG}.

    \item[EG4 Modelling Layer: Business Models on Grid] ~

The market enabled Grid infrastructure gives possibilities for new business
 models and \emph{E-Commerce}. Implementation and any significant work on Layer
 EG4 is not done until now. No open real markets exist for virtual resources.
 The lack of rentability and real business use-cases block further developments.
 Nonetheless, there are some projects such as GridEcon \cite{GridEcon},
 BIS-GRID (2007-2010) \cite{Bisgrid} and NESSIGrid \cite{Nessigrid}, which have vowed to work on business models
 for the Grid. The BIG \cite{BIGproject} project has also worked on this layer.

\end{description}

\begin{figure}
  \centering
  \includegraphics[angle=0,width=0.8\columnwidth]{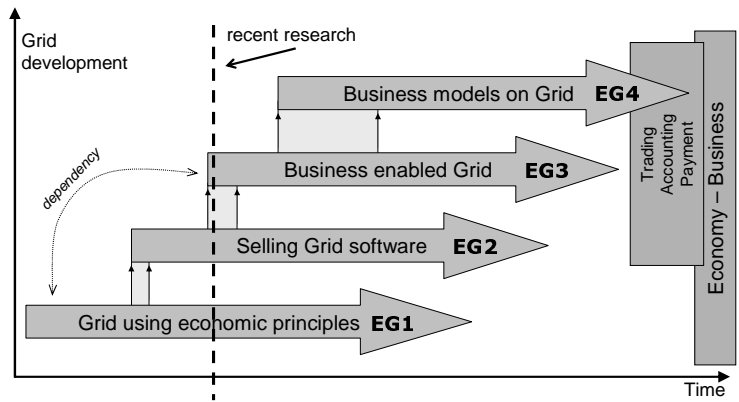}\\
  \caption{Economy-Grid layer (EG layer) model}\label{egl-fig:eg-layers}
\end{figure}

Figure \ref{egl-fig:eg-layers} shows the context of the EG-layer model with Grid development progress, time and economic usage. The recent research work is done on the first three layers only. Higher layer depends on lower layer, whereas the width of the light-gray vertical bars represents the quantity of dependency. Layer 1 interacts with Layer 3, because e.g. about a price of a resource of Layer 3 has to be agreed by a resource allocation mechanism (broker, scheduler) of Layer 1. All layers are necessary to establish the economic usage of a future Grid. The difference between Layer 3 and Layer 4 is, that Layer 3 infrastructure provides business between known partners. The Layer 4 infrastructure establishes sophisticated business models and highly dynamic virtual organizations. A business model describes the operations of a business including the components of the business, the functions of the business, and the revenues and expenses that the business generates.

\section{Dictionary on Business in the Grid}
\label{sec:dictionary}

In order to provide an overview about different Grid and business
activities, we have created a dictionary. It was written on a Web
platform using a special WIKI technology, which is particular
advantageous for a dictionary because of the many cross links that are
inserted automatically.

In order to keep it concise, the items constituting the dictionary
  were divided into five sections: \textit{Projects},
  \textit{Conferences}, \textit{Companies}, \textit{Products}, and \textit{Dictionary} that constitutes the core part. In the following the most
  interesting facts and the main results of each section are
  described.

\subsection{Projects}

There are already many projects in the area of Grid computing not only dealing with
 technical issues but also with issues concerning business aspects of the Grid.
 The Information Society Technology (IST) priority program from the European Union - one of
  the priorities of both the fifth (1998-2002) and the sixth (2002-2006) Framework
  Programmes (FP5, FP6) from the European Union - has a sub-unit called Grid Technologies
  \cite{ISTGrids} that comprises EU-funded RTD activities in the area of `Grid-based systems
  for Complex Problem Solving'. This topic been chosen as one of the `Strategic Objectives'
  of IST under the specific program `Integrating and Strengthening the European Research
   Area' of FP6. Some of the Grid-related projects deal explicitly with business aspects
   of the Grid, as for instance the GRIA project \cite{GRIA} under FP5
    and InteliGrid \cite{InteliGrid}, NextGRID \cite{NextGrid} or Trustcom \cite{Trustcom}
    under FP6. And of course there are several international projects focusing on the
    same aspects, as for instance the Japanese BizGrid project \cite{BizGrid}, the
    Australian projects Economy Grid \cite{EconomyGrid} and Gridbus \cite{Gridbus}, the
    UK Grid Markets Project \cite{UKMarkes} or the Grid Economics project
     \cite{GridEconomics} hosted at the Zurich IBM Lab.

Most of the projects have partners both from academia and industry. Some of them produce
output that is directly used by the industrial partners. This is for instance the case in
the Corporate Ontology Grid (COG) project \cite{COG}: the results obtained during the
project were directly applied to the automotive industry. As in numerous other projects,
one of the goals was to demonstrate the applicability of Grid technologies to industry.

The projects can roughly be classified in two categories: there are software-centred
projects and architecture-centred projects. The software-centred projects can mainly be
 located on the EG1 - Integration layer of the Economy-Grid layer model
 \cite{WeisSchiGridEco04} presented earlier. The outputs of these projects are mainly
 Grid-related software packages that use economic principles, such as accounting systems
  and implementations of negotiation models between the resources. As examples we can
  mention the SWEGRID project \cite{SWEGRID} and the Gridbus project \cite{Gridbus}. The
   SGAS (Swegrid Accounting System) developed within the SWEGRID project is a Java
   implementation of a Grid Accounting System based on the OGF Open Grid Service
   Architecture. The middleware developed within the Gridbus project is engaged
    to support e-science and e-business applications. Among other features it includes a
    competitive economy-based Grid scheduler, a Web-services based Grid market directory and
    Grid accounting services. Architecture-centred projects are on a more abstract
     basis (but of course some of them do also develop some software) and can mainly be
     located on the EG3 - Enabling layer of the Economy-Grid layer model
     \cite{WeisSchiGridEco04}, for instance the GRIA project \cite{GRIA}. GRIA is a
      Grid aimed at business users and enables commercial use of the Grid in a secure,
      interoperable and flexible manner. The goals are to allow industrial users to trade
      computational resources on a commercial basis to meet their needs more
      cost-effectively and to allow service providers to rent out spare CPU cycles and
       thus      to allow clients to hire those CPU cycles. In the longer term, the project
        partners aim to use GRIA as a vehicle for prototyping new business models on
        the Grid.

There are several issues appearing in mostly all projects. The most important one seems to be security. In fact, one big challenge in moving from academic Grid prototypes to enterprise Grids is the need for a high security level in the enterprise environment. This issue becomes even more important when the Grid infrastructure jumps out of the boundaries of a single enterprise. Moreover, nearly all projects are committed to contribute to the development of standards, which are a main requirement of the industry to assure interoperability among various systems. Another important point is simplicity: in the community there seems to be the conviction that only relatively simple systems can survive in the market.

Until now, there have been only very few projects that could be located on the EG4 - Modelling layer of the Economy-Grid layer model \cite{WeisSchiGridEco04}, and much work needs to be done at this level to make the economy prepared for the rapidly evolving Grid infrastructure and many possibilities within it. Until now, there seems to be a lack of clear definitions and market-regulating mechanisms, as for instance pricing models.

\subsection{Conferences, Workshops, Alliances}

Using the Grid for developing new business models is still a rather new idea, but there are already some organizations putting a lot of effort in this area. The two key organizations were the Global Grid Forum and the Enterprise Grid Alliance \cite{EGA}; the most relevant conference is the GridWorld conference that took place for the first time in autumn 2005. The Global Grid Forum (GGF) and the Enterprise Grid Alliance (EGA) have merged together on June 26, 2006 to form Open Grid Forum (OGF).

\subsection{Companies}

During the past five years, many companies providing Grid or Grid-related products and services appeared on the market, often as spin-offs from universities. We tried to classify these companies following the Economy-Grid layer model \cite{WeisSchiGridEco04} presented earlier in this chapter. It appears that most of the companies can be located on the EG2 - Commercialisation layer: they are mostly selling Grid software. As examples, we can cite Avaki \cite{Avaki}, Axceleon \cite{Axceleon}, GridSystems \cite{GridSystems} or United Devices, and  also much larger and established companies such as Oracle and IBM. However, some companies are getting a step further and can already be located on the EG3 - Enabling layer, as Parabon Inc. \cite{Parabon} for instance. Parabon provides a compute engine for free download and thus allows everybody to become a paid provider. Jobs submitted by paying clients are split in small tasks and distributed among the providers that perform them in their computer's idle time, without even noticing it. The results are then reassembled and sent back to the client. The price calculation is among other things based on task prioritization: the more quickly a client wants his job to be executed, the more he has to pay. However, most of the time the companies implementing such a system also sell some software components and it is not clear to what extent the usage of a real Grid infrastructure for doing business contributes to the overall revenue.

\subsection{Products}

As already mentioned in the section about the projects, many
Grid-middleware products have been released. Several of these products
implement economic principles such as accounting systems, negotiation
mechanisms, etc. The best known Grid middleware software is still the
Globus Toolkit now available in version 4 \cite{FosGT4}; it is used in
many projects as a basis for setting up a Grid infrastructure. The
Globus Toolkit developed by the Globus Alliance \cite{Globus} is an
open source software toolkit used for building Grids. Another example
of a Grid middleware is the open-source software developed by the Open
Middleware Infrastructure Institute (OMII) \cite{OMI}. Another
important and widely-used software is gLite
(http://www.gLite.org). Besides theses open-source products developed
by research communities there is of course a wide range of commercial
Grid products developed by companies focusing on Grid technologies, as
mentioned in the section about the companies.

\subsection{Concepts}

We summarized different topics and concepts concerning business in the
Grid and not fitting in any of the other categories. These topics
include architectural issues such as Service Oriented Architecture and
the Open Grid Service Architecture (OGSA), concepts and
technologies enabling business in the Grid, such as Virtualization,
Web Services and Grid Services, models such as the Economy-Grid layer
model \cite{WeisSchiGridEco04} and the Enterprise Grid Alliance
Reference Model \cite{EGAReferenceModel}, standardisation issues and
concrete standards such as WS-Agreement, and topics of general
interest for business in the Grid such as Grid Security, Grid Pricing
Models, Grid Accounting, etc. Some of these issues are already mature,
as Grid security for instance: because of the strong need for a secure
environment in enterprise Grids, much effort has been put on security
issues and technology seems to be very advanced in this area. On the
other hand, some of the mentioned topics are only at the very
beginning of the development. This is for instance the case with the
pricing models in the context of business Grids. There is not much
literature on this topic yet and there are no concrete proposals on
how to regulate the pricing for resources in a commercial Grid
environment.

A software licensing model is one of the things often discussed when
using commercial software on the Grid but no consensus has been
reached on that topic. However, it is clearly visible that the commercial usage of Grids or more generally speaking  business in the Grid is driving forward. More and more companies are adopting Grid technologies both inside their business and for cooperating with partners. One indication for this development is the Oracle Grid Index Report \cite{Quocirca_Report-2nd} performed by the Oracle Corporation  and the IT analyst company Quocirca. The report aims at analysing companies' attitudes towards the adoption of Grid-related technologies.

In conclusion, we can say that there are different levels of doing
business in the Grid - or different ways to interpret the expression
``business in the Grid''. These levels are conforming to the layers of
the Economy-Grid layer model \cite{WeisSchiGridEco04} developed within
the BIG project. The dictionary gives an overview of projects,
alliances, products, companies and concepts located on any of these
levels. The challenge about the Grid and especially about doing
business in the Grid infrastructure is not so much technical but to
make work together people from IT management, computer scientists,
other scientists and engineers, vendors and people from the commercial
world, that have all different visions and expectations towards the
Grid.

\section{Market Analysis}
\label{chap:market}

Several Grid initiatives deal with preliminary business
applications of the Grid (see also http://www.beingrid.com). Often,
non-Grid solutions are very particular to certain domains such as
e-business etc. In the BIG project we focused on the basic questions
of:

\begin{itemize}

\item Is there really a commercial market for the Grid?

\item How do business and IT leaders think about (adopting) Grid
  technologies?

\end{itemize}

Many people in the Grid community have their personal opinions on
these questions, and they are very diverse from being very positive to very
negative. However, we took the approach that we directly contacted
several national (Austrian) and multi-national companies in order to
{\bf interview them face-to-face} about the market potentials of
Grids. In total, we interviewed more than 25 key persons from different
companies and organizations, in order to get an idea about the current
perception of Grid technologies in business and industry. Interview
partners were mainly the company leaders (CEOs), IT leaders (CIOs, IT
managers and IT project managers) as well as technology leaders and
people involved in education. Therefore, we gained a good overview
about current strategic decisions of companies with respect to Grid
computing. Details on the interview series are given in
Section~\ref{sec:interviews}.

\begin{figure}[htb]
  \centering
  \includegraphics[width=0.8\columnwidth]{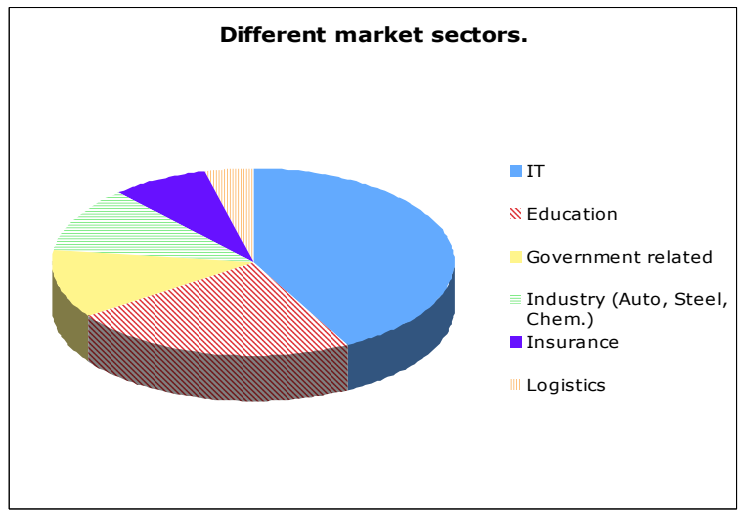}
  \caption{Different market sectors represented in the interview series. Some companies or representatives operate in two sectors (e.g. IT and Education)}
\label{fig:sectors}
\end{figure}

We also started a second survey that potentially reaches a bigger
community of companies since the {\bf interview forms are on-line}
and no direct interviews are necessary. This interview series was
done in co-operation with the CAS Community~\cite{CAScommunity},
that helped us promote the Web site to a large number of companies
world-wide, mainly in the storage business. A detailed discussion on
the on-line survey is given in Section~\ref{sec:survey}.

As a short summary, we can deduce the following:

\begin{itemize}

\item Only a rather small number of companies are currently {\bf actively
  deploying Grids} or related technologies. Often, these are
  multi-national companies.

\item For small and medium companies (SME), the idea of {\bf Grid computing
  is not yet discussed much} since even technologies such as Web
  services are yet not  fully adopted.

\item All companies see Grid computing as a way to reduce costs in
  certain areas, either for providing new services to customers
  (although such services are currently very rare) or by
  {\bf outsourcing} customer related services or computing services that are
  related to the production chain of companies. However, outsourcing
  is considered to be more attractive for bigger rather than smaller companies.

\item Some companies (such as ASPs) state that they make more money
  with {\bf selling licenses} rather than selling services.

\end{itemize}

In the following two sections we go into the details of the personal
interviews and the on-line survey.

%
%
\subsection{Personal Interviews}
\label{sec:interviews}

The usage of Grid technology in an industrial environment or within a
company can be regarded as a strategic decision such as the adoption
of a company wide Intranet infrastructure. Consequently, in order to
get Grid technologies accepted within such domains, it is important to
first ``convince'' representatives of strategic management or opinion
leaders and then people at the operational level. We therefore created
an interview series that was particularly focused on opinion leaders.

The main idea was to personally talk to people who represent companies
of different business sectors (see Figure \ref{fig:sectors}). We either contacted people directly via
e-mail or phone, or filled in requests in company Web sites that then
referred us to the respective people. In general, about 70\% of the
contacted companies or representatives welcomed us for interviews.

Most of interviewed people were directly interviewed at the company
location (office, meeting rooms). A smaller number of interviewees also
visited us at the university for an interview.

In order to prepare for the interview and to structure the questions
to ask we created a questionnaire that contained
several questions, focusing on the following main areas:

\begin{itemize}

\item {\bf General Questions} about the company: this section of the
  questionnaire contains the sector in which the company is acting, the
  size, age of the company etc.

\item {\bf The Interviewee} (person to be interviewed): Here, we asked
  the interviewed person to give some public information about his/her
  position within the company.

\item {\bf IT Questions}: Since Grid computing is not yet a term that
  all opinion leaders are fully aware of, we first wanted to ask some
  general IT questions such as where IT is used? Does the company use
  outsourcing, networking etc.? Are there storage and/or computing
  intensive demands? These questions help focus and concentrate on
  features that Grid computing mainly provides.

\item {\bf Grid Questions}: The main part of the survey then focused
  on Grid questions, common understanding and the application of Grid
  technology within the company or in relation to customers and their
  products. We also analysed if a service oriented software
  architecture is used.

\end{itemize}

Typically, one or two representatives of the BIG project
were leading the interviews. Each interview lasted between 30 and 60
minutes. The list of interviewed companies (which were located in Austria, Germany,
Slovakia, France, and USA) is the following:

Microsoft Austria, SAP Germany, Oracle Austria, Sun Microsystems Austria, Bull Austria, Apple Computer Europe, Telekom Austria, Magna Steyr, WIFI Wien, MCNC, BMVIT, Uni Wien, BeoC, EC3, 1H3G (Drei Austria), IP Center, Muehlehner \& Tavolto GmbH, Uniqa, Tecco, TNT, Voest Alpine, BASF, Allianz, Austrian Social Insurance Authority for Business (Sozialversicherungsanstalt, SVA), The Austrian Federal Economic Chamber (Wirtschaftskammer Oesterreich)

The detailed results of the interviews are confidential and elaborated
in the following section. However, after the interview series several
{\bf common statements} and {\bf observations} were made that we can
summarise as follows:

\begin{itemize}

\item A good {\bf cost model} is required. The adoption of any
 new (computing) technology costs money since old systems have to be
 replaced by new ones. In addition, Grid technology typically
 promises that more storage and computing power is available to process
 faster or allow for higher data availability. However, companies need
 a simple cost model which can be used to express what is the real
 cost cut by usage of Grid technologies. Simply put, how much
 money can a company save in which amount of time?

\item A {\bf good accounting/billing} is needed: Once Grid technology is
  adopted, companies want to make money with the usage of Grid
  resources. Similar to the cost model, a clear and precise accounting
  system is need that records the usage of customers and charges them
  accordingly.

\item One of the main ideas of Grid computing is the {\bf sharing of
  resources}. However, in a business environment sharing of data
  creates conflicts such as ``who owns that data?''. Some people
  show no willingness of sharing data or even giving it away to be stored in
  remote places outside the administrative domain of a certain company
  branch.

\item Grid computing is often still considered to ``just'' provide
  either computing power and/or storage. Several companies such as
  Sun, Amazon
  or Bull provide pure computing and storage on demand. However,
  customers need services rather than just pure CPU cycles or
  storage. Therefore, \textbf{availability of high-level, application oriented
  services} is very important in the commercial field.

\item {\bf Service Level Agreement} is required for all the services
  offered by the Grid. This basically means that end users get a
  guarantee about reliable request execution within some given
  time. Furthermore, results that are obtained by applications being
  executed in the Grid need to be verified.

\item {\bf Network bandwidth}. Although the effective bandwidth of
  wide area network links is constantly increasing, the costs for
  commercial usage are still considered rather high by certain
  companies. Therefore, several business customers will benefit more
  from Grid computing once network prices will further decrease.

\item {\bf Technology adoption takes time}: Grid computing is
  currently very much related to Web services and a service-oriented
  architecture. Many SMEs still have not even adopted .NET or
  similar Web services technologies such as SOAP etc. Therefore,
  several companies might remain conservative and will not adopt Web
  services in the near future.

\item {\bf ASP profitability}: Until now many CEOs in companies believe
  that selling software is more profitable than selling services. The business
  case of earning money by application service provisioning is (at least in Europe)
  not that accepted as an opportunity. However, most of the interview partners
  acknowledged that ASP will be a big issue in the near future
  (for SMEs as well as for industry).

\end{itemize}

\subsection{On-line Survey}
\label{sec:survey}

The second survey carried out during the project, was mainly addressed to the people already familiar with Grid computing and working with it. Similar to the questionnaire developed for the face-to-face interviews, interview forms were created but this time the were made available on-line.

Therefore, no direct interviews were necessary. The survey has been
promoted by the CAS Community (http://www.cascommunity.org), a community working in the area of storage solutions. Thanks to this cooperation, a larger number of companies could be reached. The survey has also been publicized at the 2005 Grid World Conference in Boston, MA, USA, where we could reach many Grid specialists.

The surveys have been filled out by people both from industry and academia. Most of the represented universities and companies are rather big institutions, with several thousand employees and internationally distributed branches. Most of the companies have the emphasis on software sales, and they are all interested to investigate new information technologies.

The interviewees were mostly sales consultants, system managers and researchers, with at least a few years of working experience and a solid knowledge about Grid computing. Most of them have been knowing about Grid technologies for 4 to 5 years already.

As a summary, we can deduce the following most important observations for the on-line survey:

\begin{itemize}
  \item The term ``Grid computing'' seems still not to be well defined, as people have different opinions on what Grid computing exactly means. Most commonly, Grid computing is associated with ``Distributed computing'' and ``Virtualization of IT resources''. Many also refer to Grid computing as an ``infrastructure that allows an efficient sharing of data across different locations''. Only some of them quoted that Grid computing is a confusing term with different meanings. Only one person answered that the Grid is an ``infrastructure that opens new ways of doing business'', from which we can deduce that - as assumed - the Grid is still associated with technical issues rather than economical ones. Also, the visions companies have about the Grid, are mainly application-focused. The Grid is seen as a means to better achieve goals in specific disciplines, e.g. in biotechnology.

  \item For the companies that completed the survey, Grid computing is an important issue or even the main business area. It seems that those companies interested in Grid technologies put a lot of time and effort in it - and do not just run a Grid in order to experiment with it. Most of the companies are doing ``Business in the Grid'' by selling Grid software. They are thus on the ``EG2'' of the Economy-Grid Layer Model~\cite{WeisSchiGridEco04}. But some quoted that they were doing Business in the Grid by ``using a Grid infrastructure to provide resources and/or services'', which could place them on the 3rd layer of the Economy-Grid Layer Model.

  \item Conforming to the companies and research institutions, Grid technologies will continue to establish themselves on the market and even become more important. They are said to become more important on the technical level, and 85\% of the interviewees think that Grid technologies could have an impact on business models. We can thus deduce that there is an industry awareness concerning Grid and business. The results seem to underline the estimations found in recent papers related to Grid computing: until now the technical challenges prevail, but there is awareness for the possible economical applications of Grid computing.

  \item Grid solutions seem to be applied to many different business areas (such as Customer Management, Product Development, Supply Chain Management) and across all industries. About 70\% of those who have completed the survey provide IT resources or services on demand over the Grid infrastructure, and there are already Grid solutions supporting economic principles, such as accounting functionalities. This result is rather surprising, as we thought that the implementation of economic principles in Grid solutions is not yet widespread.

  \item Finally, all the interviewed companies support research on Grid technologies - either in-house, or by providing funding, or both. It is also worth mentioning that many companies claim to have academic partners that do research on Grid technologies, and to be in touch with international Grid organizations or communities such as the Global Grid Forum.

\end{itemize}


\subsection{Web Portal}\label{sec:portal}

Another outcome of the project is a Web portal for general information about business Grids. It can be used for retrieving information collected in the project in general, especially for the parts ''Revisiting of Existing Business Models'' and ''Market Analysis''.

The portal is based on CEWebS (Cooperative Environment Web-Services) \cite{Mangler2005}. This environment provides:

\begin{itemize}
  \item The so-called Scientiki, which is a WIKI that has the additional capabilities of exporting to \LaTeX{} and HTML, and thus supports the creation of scientific papers.
  \item An evaluation part that handles forms, questionnaires and surveys.
  \item The possibility to run a discussion forum. CEWebS also provides other functionalities, but they were not used in this work.
\end{itemize}

\begin{figure}
\begin{center}
  \includegraphics[angle=270,width=0.8\columnwidth]{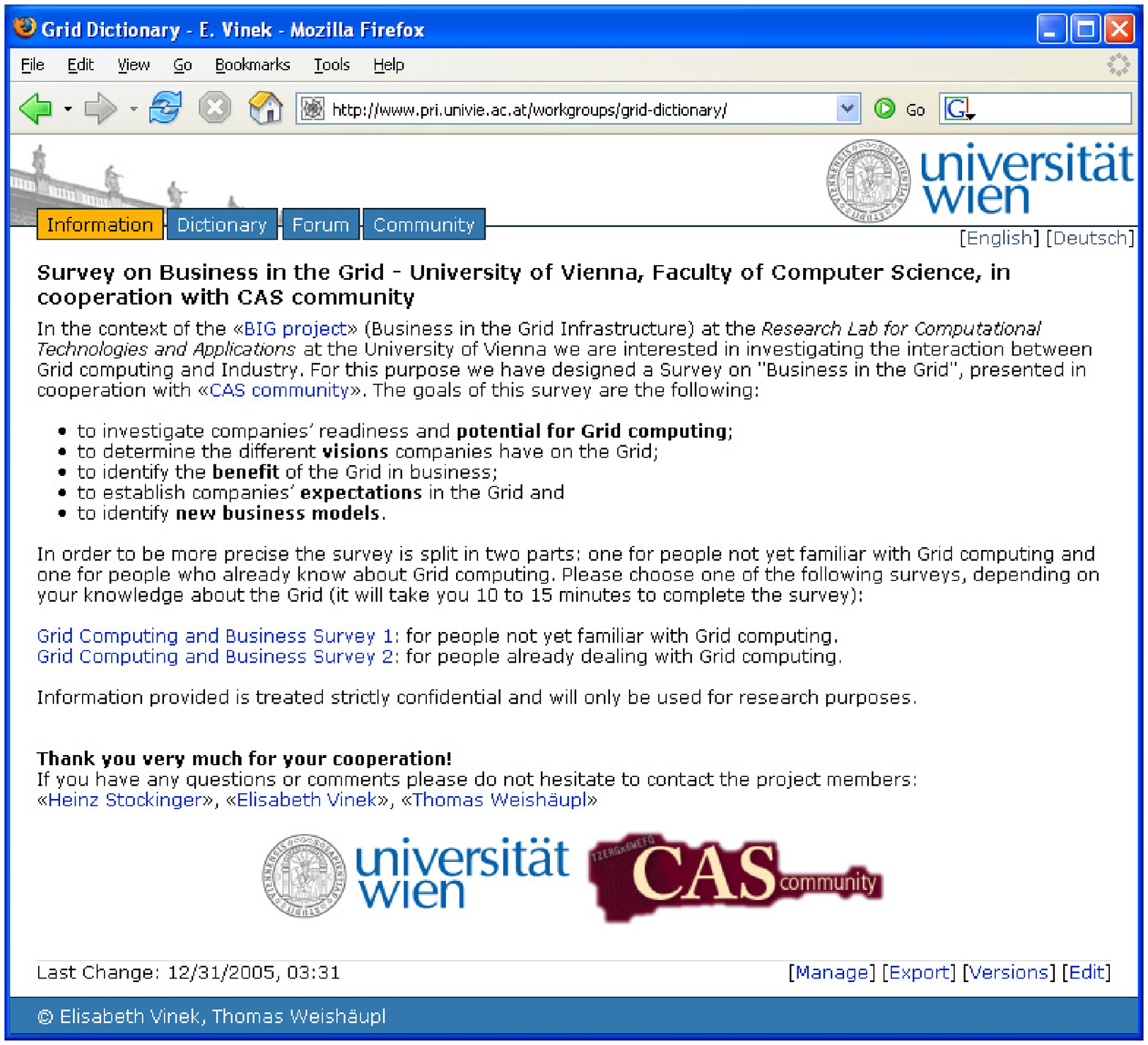}
  \caption{Web Portal}\label{fig:WebPortal}
\end{center}
\end{figure}

The portal is available at the address \texttt{http://www.cs.univie.ac.at/big} and contains the following four parts:

\begin{itemize}
  \item \textbf{Information \& Surveys}: The start page of the portal contains a short project description and links to the online surveys mentioned in the Market Analysis in section \ref{chap:market}. The start page and the surveys are available both in English and German. As already mentioned, the surveys were presented in cooperation with the CAS Community \cite{CAScommunity} who helped promote the surveys by announcing them in their newsletter and on their Web site. Figure \ref{fig:WebPortal} shows the start page of the Web portal.
\end{itemize}

\begin{itemize}

  \item \textbf{Dictionary}: This dictionary is the result of a master thesis ``Dictionary on Business in the Grid'' \cite{Vinek2005} realised within the project . In the Web portal, all the items contained in the dictionary are listed and can be clicked. They are partitioned into the following categories:

  \begin{itemize}
    \item Introduction: contains general information and definitions about the Grid and Business in the Grid.
    \item Projects: here several projects dealing with Business in the Grid are described.
    \item Conferences, Workshops, Alliances: this section contains all the conferences, workshops and alliances that are considered relevant for Business in the Grid.
    \item Companies: description of several companies providing Grid or Grid-related solutions and services.
    \item Products: here Grid products that contain some business aspects are presented.
    \item Concepts: this section contains the core of the dictionary. As in the other categories, the items are listed alphabetically.
    \item Links: here general interesting links to Grid resources can be found.
    \item Conclusion and Outlook
  \end{itemize}
\end{itemize}

\begin{itemize}
  \item \textbf{Forum}: As an additional functionality, a discussion forum is available. It is meant for communication both among the project members and with external people interested in the project. The forum supports RSS integration - users can thus subscribe to the forum with a mail program and receive the messages as e-mails.
\end{itemize}

\begin{itemize}
  \item \textbf{Community}: In this part the project members are listed with their e-mail address so that interested people can easily contact them.
\end{itemize}

\section{A New Business Model}
\label{chap:new-models}


In the following chapter we illustrate one key business model
that was identified and elaborated within the BIG project which builds an enabling basis for
implementing the vision of selling software and resources as a service
and not as a good. It is based on ``Trust and Security''
(Section~\ref{sec:gset}) and also resulted in a prototype based on
state-of-the-art Grid technology.

\subsection{Trust and Security: gSET}\label{sec:gset}

  Trust and security are often claimed in Grid computing as some of the
    functional differences to earlier developments in the Web and
    distributed computing \cite{Fo02b}. Beyond organizational
    boundaries, virtual organizations need a trustable and secure
    infrastructure to utilize autonomic resources and services. Security
    describes a field of activities to guarantee the privacy, integrity
    and availability of resources.


  \subsubsection{The Need for Dynamic Authorization} 

  Although Grid research has gone a long way from the first steps in the 1980s,
  it is still far from providing plug and play tools that can easily be deployed
  and maintained. Current Grid Solutions are mostly custom fitted systems deployed by big players
  like IBM for big customers or installed at scientific research facilities fitted to the needs of scientists
  that are often very different from the requirements a business poses on a new technology.

  Furthermore, most Grid projects are still very introspective. They are intended to share the resources of some high
  performance computing center or provide a common file system to members of a research group. They focus on strongly coupled
  rather small virtual organizations while the vast majority of the benefits Grid computing could leverage keep lying bare in
  the Internet.

  Publicly available Grid services however will serve a heterogeneous group of customers not bound together by being part
  of the same business or working at the same scientific working group. This demands for tools to dynamically manage
  big communities of loosely coupled entities providing and consuming Grid services without excessive administrative overhead.

  Another problem hindering the progress in publicly available Grid services is the method of payment.
  Conventional services like delivering pizza or books or booking a flight that are offered over the Web require human
  interaction anyway. Therefore payment can be handled interactively by the requester of the service. However Grid services
  typically only involve machine to machine interaction and should be processed transparently for the user. Today there are
  very few projects that address this problem.

  One of the most promising usages of Grid computing is distributed storage and manipulation of data. For a flexible and
  universal usage of Grid resources, it will be necessary to abstract the process of distribution and make it transparent for
  the Grid user.

    A metaphor often used for this abstraction is the electric power Grid. What seems to the user as a simple push on a button
    is backed by a sophisticated framework of power plants and transformer stations that span (sometimes) the whole continent.

    To obtain this abstraction it will be crucial to have an opportunity to provide a flexible way to manage the access to Grid
    resources anonymously as well as trusted.

    While the provider of a resource wants to secure that the user of his facilities is trustable, the user himself is interested
    in maintaining his anonymity, while gaining easy access to the services.

    Staying with the metaphor of the electric power Grid, the user does not want to have to identify himself to an operator of
    the electricity provider if he turns on the light (and therefore uses their services). In turn the electricity corporation
    does not want to face the bureaucracy arising with identification for every watt they send him. Instead they are interested
    in knowing him as a reliable customer with a sufficiently covered bank account to pay for his demand of electricity.

    In the course of the BIG project we developed gSET \cite{WeisWit2006}, a mapping of Secure Electronic Transaction (SET) \cite{SET_1,SET_2,SET_3} to the Grid environment, which can fulfill
    these requirements.

    It is intended to show a possibility to handle authorization and payment in loosely coupled Grids of commercially
    interacting units. Using Grid services should be as easy as paying with a credit card (or even easier).

    The method described provides an authorization mechanism that is \emph{request based} and \emph{dynamic}.

    \emph{Request based} means that the authorization to use a service is evaluated at the time the service is to be invoked
    and can depend also on the contents of the service invocation not only on the role and rights of the service requester.

    Furthermore in gSET the authorization to use a service depends on the service requester's credit at a configurable account
    provider, making the solution highly \emph{dynamic} and easy to integrate into existing accounting systems.
    Due to this delegation a gSET enabled service does not require an extensive account management by the service provider and is
    therefore perfectly suitable for the commercial service provision.


    \subsubsection{Requirements for Dynamic Authorization}

        gSET aims at providing a framework that enables a provider of a Grid service
    to make his service accessible for everyone willing to pay
    the price set by him.

    gSET must address the following important issues:

    \begin{itemize}
        \item Relieve the service provider of complex account management and payment issues
        \item Secure the transaction against observing and manipulating attacks
        \item Protect the service requester's confidential data against fraudulent usage
    \end{itemize}

    To relieve the service  provider of account management and payment issues while guaranteeing the trustability of the service
    requesters, we have to delegate these topics to a third party. Like in current credit card payment solutions, we need a payment
    service that evaluates the credit of the service requester and manages the transfer of credits.

    To secure the transaction against observing and manipulating attacks we must use a combination of digital signatures and
    encryption. Fortunately Grid environments always contain public key infrastructure so we can assume each participant owns an
    identity certificate along with a public/private key pair.

    The protection of the service requester's confidential data against fraudulent usage has two aspects. Firstly we do not want
    the service provider to know payment related information like the account number or a possible password. Secondly the
    payment service should not receive information about the nature and the parameters of the consumed service.

\begin{figure}
      \centering\includegraphics[width=0.6\columnwidth]{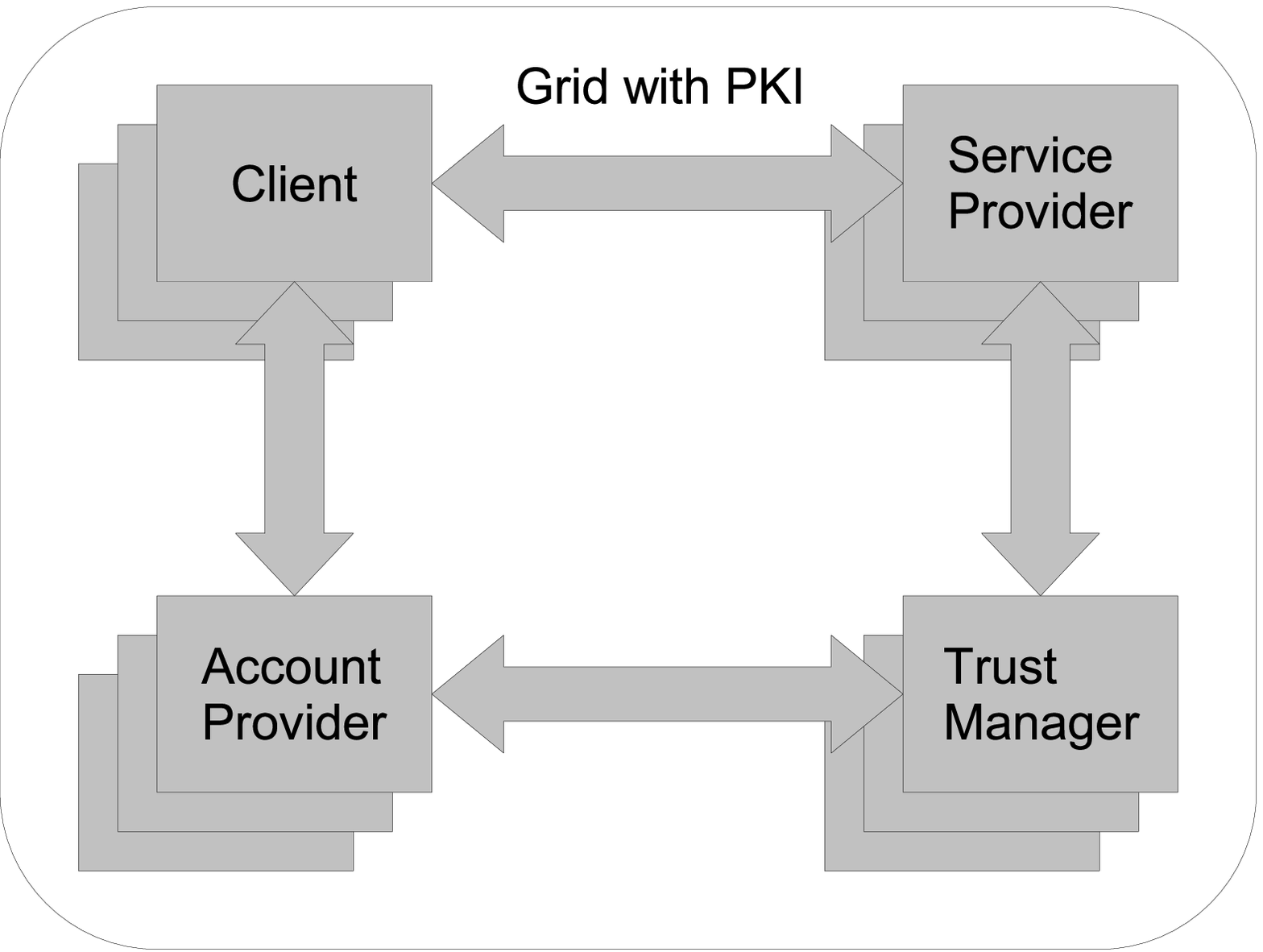}
       \caption{gSET architecture}\label{fig:gsetarch}
    \end{figure}

    We can now identify three actors in our scenario, the \emph{service provider}, the \emph{service requester} and the
    \emph{payment service}.
    If we look at current electronic payment solutions however, we find that the \emph{payment service}
    is in most cases split up into two closely related but nevertheless independent units. In our case we will call them
    \emph{account provider} and \emph{trust manager}. While the \emph{account provider} maintains the account and the
    relationship to its account owners, the \emph{trust manager} is responsible for the technical execution of the payment
    process. The architecture of the gSET actors is depicted by Figure \ref{fig:gsetarch}.

    If we take a closer look at the situation we can identify five use cases.

    \begin{itemize}
        \item The Service Provider contracts a Trust Manager to handle the authorization and billing.
        \item The Service Requester contracts an Account Provider to vouch for him.
        \item The Service Requester requests the price for the Service.
        \item The Service Requester uses the Service.
        \item The Service Provider collects his credits.
    \end{itemize}

    With gSET the latter three use cases are addressed, while we assume that the first two
    are already successfully provided by the Grid infrastructure.

    Before the service requester uses the service he must have the possibility to retrieve the price the service provider
    will charge him for the requested operation.
    To get this information the service requester must send information on how he wants to use the service.

    When the service requester decides to use the service, he must first request authorization. To obtain authorization
    he must send the information on how he wants to use the service along with which account provider he wants
    to use, his account details, and the amount the trust manager should be allowed to authorize.
    It must be guaranteed, however, that the account details and the amount to be authorized are not visible to the service
    provider,   nor can be manipulated by him.

    The service provider relays the account details and the amount to authorize (but not the information on how the service
    requester wants to use the service) to a trust manager that can manage transactions for the service requester's account
    provider along with the amount he intends to charge the service requester.
    The trust manager checks if the charged amount lies within the limit given by the service requester and the credit the
    service requester is granted by the account provider. If both checks succeed he stores the information on the amount the
    service provider charged and sends a positive reply to the service provider along with some kind of token that the service provider
    can use to collect his credits later.
    If the service provider receives a positive reply he allows the service requester to use the service in the desired way.


  \subsubsection{A Use-Case for gSET: Storing Data in the Mobile Grid}

  Let us visualize this scenario: Jim is an ambitious tourist who wants to record his memories by taking photographs of all the historical
places of Vienna by using his 3.2 Mega Pixel camera built into his cellular phone. But
soon after a few pictures, he realizes that he has run out of his mobile phone
memory. He suddenly recalls the Grid services accessible by a broker. He seeks the
address of the broker from his bookmarks in his mobile and sends it a request for
storage of pictures that he has already taken. The broker responds promptly and
sends him links of different Service Providers and their rates. When Jim selects a
Service Provider, the Service Provider responds back and the payment process uses
Jim's credit card. Jim's credit card information is already saved in his mobile
device but he is confident that this is a very safe process and there is no need to
worry about anything. He realizes that the data has started migrating, but what a
pleasant surprise! He notices that the mobile device is using Bluetooth to move the
data, thus no fee will be charged for the transmission. After the successful
migration he receives some thumbnails of the images which are in fact tickets that
he can use later on to retrieve his pictures. He wants to send some pictures, that he
has already taken, to his friends. He instead sends some tickets of the pictures
through SMS. After a couple of days he uses the rest of the tickets to retrieve his
data.

In the above architecture we make it possible to sell and buy Grid services through
cell phones. Our business model is based on simple and straight forward principles.
\begin{itemize}
        \item The customer can buy things without revealing her/his financial information to the
vendor.
         \item Customer and vendor both express their trust in a mediator who assures
the customer about the reliability of the vendor and guarantees the vendor about
the financial status of the customer.
 \end{itemize}

 We propose a novel business model through our state of the art financial
mechanism called gSET . gSET keeps the financial information separate form the order
information and enables both parts of the information to reach the appropriate
players i.e. the Trust Manager and the Service Provider respectively. The Trust
Manager plays a very important role in a financial transaction in the sense that both
the Service Provider and the Account Provider have trust in it. The Account Provider
verifies the client's financial status and thus the Service Provider has no
hesitation to process the order. In the scenario that we have presented here, the
customer buys storage services from a vendor. Customer's financial credibility is
maintained with the Account Provider that may be at the same time the mobile
service provider and charges the customer on a regular basis. The Account Provider
pays for the service on behalf of the customer and this amount is added to the
monthly bill of the customer. Account providers need to have a contract with the
Trust Manager. The broker can offer cheap and premium accounts to the Service
Providers and charge them on a regular basis. Customers can rank Service Providers
according to service reliability. This ranking can also be done based on sales
patterns. A more reliable Service Provider gets a higher position on the list of
Service Providers that is sent to the customer by the broker.

  The gSET model highlights five dimensions of doing business on the Grid.
    \begin{itemize}
        \item \textbf{Transparency.}    The gSET model enhances the concept of transparency in the
        Grid. The service provider and the user are unaware of each other while
        the Trust Manager plays a mediating role and helps to establish trust
        between these strangers. There is a fool proof payment mechanism that
        fits seamlessly in the service oriented architecture.

        In order to achieve this kind of security, gSET could be of great help. While the encryption of the remotely stored data is
        not defined in gSET, it can help providing means of distributing sensitive data under relative anonymity while holding the
        key to putting it all together again. This minimizes the data to be physically protected and helps enhancing security.

        \item \textbf{Trust.}    The gSET model provides the bridge of trust between two
        completely unknown business partners i.e., the user and the service
        provider. This trust is guaranteed by a very secure PKI based
        communication mechanism having the trust manager as the focus. Both
        parties express trust in the trust manager which guarantees the
        credibility of both parties and yet keeps their confidential information
        secret.

        \item \textbf{Privacy.}    The service provider does not know the credit information of the user
        and at the same time the account provider is also unaware of the
        shopping-history of the user. In our example we have shown how to use
        the PKI infrastructure of the gSET to encrypt the information stored on
        the service provider's devices.

        With gSET, the service provider obtains only as much information about the service requester's identity as the service
        requester intends to give. However, this relative anonymity does not deprive the service provider of the trust in the service
        requester's reliability, as the trust is delegated to the trust manager.

        The trust manager does not receive information about the details of the services the service requester uses. If the service
        requester wants to put the level of privacy protection even higher he may even use more than one account provider to vouch
        for him.

        \item \textbf{Agility.}    The gSET model is a very secure, sound and yet flexible system as it can be fit within an
        already designed workflow. It acts as a loosely coupled payment module
        that takes care of financial transactions among business partners on the Grid. We have demonstrated \cite{emags07} how to fit this module into already
        defined business workflow on the Grid.

        With a natively distributed solution like gSET, it is easier to address the challenges of a mobile world.

        \item \textbf{Reliability.}    The gSET model is in fact a gridified version of
        the SET model \cite{SET_1, SET_2, SET_3}. SET was a very reliable, errorfree financial
        transaction system with disconnection and rollback facilities. SET
        enhances security and reliability by its certificate distribution. This
        distribution of certificate although being a very powerful mechanism was
        not accepted as it was realized cumbersome by the users and hence SET
        couldn't succeed. The case with Grid is different as it already
        implements security by distribution of certificates among the partners.
        Therefore gSET sounds like an ideal match for the Grid

    \end{itemize}

  The benefits of using gSET as authorization mechanism in business scale storage services are evident for both service
  provider and service requester.

  The service provider does not have to meddle with a complicated account management. He can easily delegate the issue
  of the service requester's trustability to a trust manager. This allows the access to a much bigger market, as the account
  providers can attract a lot of customers willing to express trust in trust managers because they do not wish their financial
  information to be disclosed all the times.

  The only requirement regarding certificates, gSET poses, is that both partners trust the certificate of the trust manager.
    In an economic market of service providers, the policy is quite simple. The provider sells to anybody who does not have any
    legal or rating problems. For this, the provider needs to ensure that the customer is trustable and pays for the used
    service. With gSET it is possible to transfer the permission to verify this information for one specific transaction  without disclosing
    the private information.

\section{Conclusion}
\label{chap:conclusion}

During more than two years lifetime of the BIG project, Grid standards and
technologies have improved quickly, and several new business models have been
investigated by the BIG team. Although Grid technologies are well established in
certain communities in research and academia (data and computing intensive science
fields such as physics, bioinformatics, etc.), engineering etc., the commercial
market is not yet ready to adopt Grid technologies. This has several reasons such as
evolving standards and partly immaturity of Grid solutions. We see a trend that the
Grid is still in the phase where \emph{early adopters} and \emph{expert users} are
the dominant users. However big software companies are promoting Grid technologies which is a good
sign. 
Finally we presented a specific business model for the mobile Grid, which is based on gSET allowing for 
dynamic authorization. gSET is the first enabling step to make Grids a platform for commercial
workflows.

\section*{Acknowledgement}

This research project was funded by
project number 10547 of the OeNB Anniversary Fund.

\bibliographystyle{plain}

\bibliography{literature}
\end{document}